# Prediction of structure-dependent thermal transport behavior in self-folded graphene film validated by molecular dynamics simulation

Anran Wei [a], Fenglin Guo [a, b*]

[a] Department of Engineering Mechanics, School of Naval Architecture, Ocean and Civil Engineering, Shanghai Jiao Tong University, Shanghai 200240, China

[b] State Key Laboratory of Ocean Engineering, Shanghai Jiao Tong University, Shanghai 200240, China

* Corresponding author, E-mail: flguo@sjtu.edu.cn

**Abstract**

Understanding the relationship between the microstructures and overall properties is one of the basic concerns for the material design and applications. As a ubiquitous structural configuration in nature, the folded morphology is also widely observed in graphene-based nanomaterials, namely grafold. Recently, a self-folded graphene film (SF-GF) material has been successfully fabricated by the assembly of grafolds and exhibits promising applications in thermal management. However, the dependence of thermal properties of SF-GF on the structural features of grafold has still remained unclear. We here develop an analytical model to describe the thermal transport behavior in SF-GF. Our model demonstrates the relationship between the geometry of grafolds and thermal properties of SF-GF. The predictions of temperature profile and thermal conductivity are well validated by molecular dynamics simulations. Using this model, we further study the evolution of thermal conductivity of SF-GF with the unfolding deformation during stretch. Moreover, the effect of geometrical irregularity of grafolds is uncovered. Interestingly, the predicted transport behaviors of SF-GF under stretch fit some analogous experimental observations reported in graphene-based strain sensor. Our results not only reveal the mechanisms behind some physical phenomenon in the applications of graphene-based devices, but also provide practical guidelines for the property design of SF-GF and other graphene assemblies with folded microstructure.

## 1. Introduction

The folded morphology is ubiquitous in nature as a basic structural feature over a wide range of length scale observed from paper origamis to nanomaterials. Folding a structure not only changes its geometrical shape into complicate phases but also brings new physical properties and enhance material performance in many applications. A well-known example of natural materials with folded microstructure is the spider silk composed of proteins. The folded $\beta$-sheet structure in proteins gives both high strength and ductility to the spider silk [1]. It is a universal law that the properties of materials strongly depend on the characteristics of their microstructure. Thus, understanding the relationship between the folded microstructures and their overall properties is a permanent pursuit for the design and application of folded architectures in multiscale.

Among various materials, graphene is one of the most famous stars due to its extraordinary mechanical [2], thermal [3] and electrical [4] properties. It can be easily warped in the out-of-plane direction to form the folded microstructure termed grafold, which widely exists in many graphene-based nanomaterials [5-7]. Theoretical and simulation studies found that the grafold can efficiently modify the electronic [8, 9] and mechanical [10] properties of graphene. Recently, a self-folded graphene film (SF-GF) material has been successfully fabricated as the macroscopic assembly of grafolds [11], as shown in Fig. 1a. The directional size control and arrangement of grafold (Fig. 1b) have been also achieved in lab [12]. It is reported that the SF-GF exhibits both good flexibility and ultrahigh thermal conductivity due to the introduction of micro-folds into graphene sheets [11]. Thus, the SF-GF is a promising material for the fabrication of next generation electronics. For example, the SF-GF has been applied on the smartphone and shows a superior performance of heat dissipation during the test [11]. However, the relationship between the geometry of grafold and thermal conductivity of SF-GF has been still unexplored till now. A comprehensive understanding of this structure-property relationship is in urgent demand for the manipulation on thermal properties of SF-GF, which will promote its industrial and commercial applications. Moreover, this relationship can support the property design of other graphene assemblies due to the widespread distribution of grafold in these material systems.

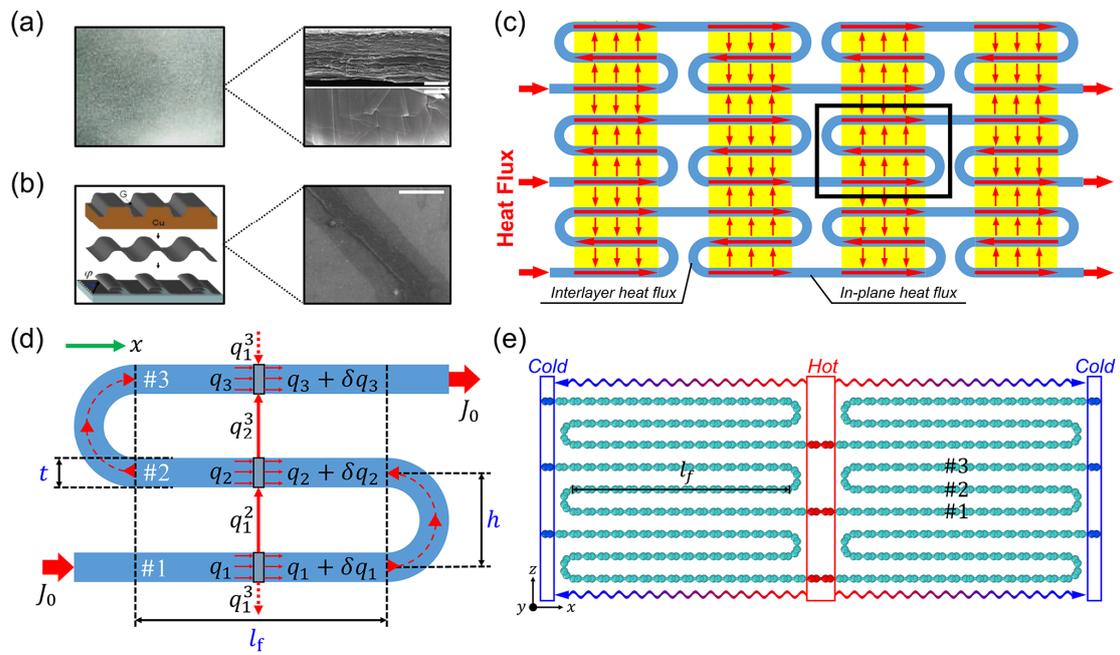

**Fig. 1.** (**a**) Self-folded graphene film (SF-GF) composed of many micro-folds of graphene as shown in the images of cross section. Scale bar: 4 μm (up) and 300 nm (down). (Adapted with permission from [11]. Copyright 2017 John Wiley and Sons). (**b**) Controllable design and fabrication of folded graphene structure with regular size and arrangement. Scale bar: 1 μm. (Adapted with permission from [12]. Copyright 2011 American Physical Society). (**c**) Structure model and the representative volume element (RVE, labelled by the black box) of SF-GF. (**d**) Illustration of heat transfer in RVE of SF-GF. Graphene layers 1, 2 and 3 are denoted by #1, #2 and #3, respectively. (**e**) Atomic model of the molecular dynamics (MD) simulation for the calculation of thermal conductivity of SF-GF.

In this paper, we here develop an analytical model to investigate the structure-dependence of thermal transport behavior in SF-GF, which is well validated by molecular dynamics (MD) simulations. Based on the model, the evolution of thermal conductivity of SF-GF under stretch is further discussed in this paper. Our model can serve as an efficient tool to predict thermal properties of SF-GF, without the restriction of computing resource for MD simulations and expensive/complicated facilities for experimental measurements.

**2. Theoretical modelling**

In our model, the SF-GF is assumed to be regularly folded with uniform fold length based on the experimental characterizations [11, 12], as the periodic model shown in Fig. 1c. The heat flow in SF-GF transfers not only along the graphene basal plane but also between the folded graphene layers due to the interlayer van der Waals interaction. Thus, the thermal properties of the SF-GF depends on both in-plane thermal conductivity of graphene ($k_\text{g}$) and interfacial thermal conductance between graphene layers ($G$). A 2D representative volume element (RVE) in the black box can be adopted to describe the overall structure. As illustrated in Fig. 1d, the RVE is a tri-folded structure with geometries of fold length $l_\text{f}$, interlayer distance $h$ and graphene layer thickness $t$. A 2D heat flow $J_0$ with a steady rate transfers across the RVE via layer 1, 2 and 3 in order as shown in Fig. 1d. Then the heat flux balance in layer 1, 2 and 3 can be expressed as

$$\begin{cases} t\dfrac{\partial q_1}{\partial x} = -q_1^{(2)} - q_1^{(3)} = G(T_2 + T_3 - 2T_1) \\ t\dfrac{\partial q_2}{\partial x} = q_1^{(2)} - q_2^{(3)} = G(T_1 + T_3 - 2T_2) \\ t\dfrac{\partial q_3}{\partial x} = q_2^{(3)} + q_1^{(3)} = G(T_1 + T_2 - 2T_3) \end{cases} \quad (1)$$

where $q_i$ ($i = 1, 2, 3$) and $T_i$ ($i = 1, 2, 3$) are the heat flux and temperature profile in layer 1, 2, 3, respectively, as well as $q_i^{(j)}$ ($i = 1, 2$ and $j = 2, 3$) stands for the interlayer heat flux from layer $i$ to $j$. Substituting the Fourier's Law into Eq. (1)

$$\begin{cases} q_1 = -k_g \dfrac{\partial T_1}{\partial x} \\ q_2 = -k_g \dfrac{\partial T_2}{\partial x} \\ q_3 = -k_g \dfrac{\partial T_3}{\partial x} \end{cases} \quad (2)$$

the governing equations for heat transfer in the RVE are written as

$$\begin{cases} \dfrac{\partial^2 T_1}{\partial x^2} = \dfrac{G}{k_g t}(2T_1 - T_2 - T_3) \\ \dfrac{\partial^2 T_2}{\partial x^2} = \dfrac{G}{k_g t}(2T_2 - T_1 - T_3) \\ \dfrac{\partial^2 T_3}{\partial x^2} = \dfrac{G}{k_g t}(2T_3 - T_1 - T_2) \end{cases} \quad (3)$$

In our model, the joint between different layers is reduced to a point at which temperature and heat flow in the neighbouring layers are continuous, since the size of joints can be ignored comparing with the large aspect ratio of folded domain. Then the boundary conditions are introduced as

$$\begin{cases} T_1(x=l_f)=T_2(x=l_f), \ T_2(x=0)=T_3(x=0) \\ \dfrac{\partial T_1(x=l_f)}{\partial x}=-\dfrac{\partial T_2(x=l_f)}{\partial x}, \ \dfrac{\partial T_2(x=0)}{\partial x}=-\dfrac{\partial T_3(x=0)}{\partial x} \\ \dfrac{\partial T_1(x=0)}{\partial x}=-\dfrac{J_0}{k_g t}, \ \dfrac{\partial T_3(x=l_f)}{\partial x}=-\dfrac{J_0}{k_g t} \end{cases} \quad (4)$$

The solutions of temperature profiles are derived as

$$\begin{cases} T_1(x)=C_1-\dfrac{J_0}{3k_g t}\left[x-\dfrac{e^{2\lambda l_f}+2e^{\lambda(2l_f-x)}+e^{\lambda(l_f+x)}-e^{\lambda(l_f-x)}-2e^{\lambda x}-1}{\lambda\left(e^{2\lambda l_f}-e^{\lambda l_f}+1\right)}\right] \\ T_2(x)=C_1-\dfrac{J_0}{3k_g t}\left[x-\dfrac{e^{2\lambda l_f}-e^{\lambda(2l_f-x)}+e^{\lambda(l_f+x)}-e^{\lambda(l_f-x)}+e^{\lambda x}-1}{\lambda\left(e^{2\lambda l_f}-e^{\lambda l_f}+1\right)}\right] \\ T_3(x)=C_1-\dfrac{J_0}{3k_g t}\left[x-\dfrac{e^{2\lambda l_f}-e^{\lambda(2l_f-x)}-2e^{\lambda(l_f+x)}+2e^{\lambda(l_f-x)}+e^{\lambda x}-1}{\lambda\left(e^{2\lambda l_f}-e^{\lambda l_f}+1\right)}\right] \end{cases} \quad (5)$$

where $\lambda=\sqrt{3G/k_g t}$ is defined as a *characteristic length scale* in the unit of m$^{-1}$. Here, $C_1$ is a parameter that is determined by the average temperature of the whole RVE and has no contribution on the effective thermal conductivity of SF-GF model.

In this model, the thickness of SF-GF perpendicular to the heat transfer direction is defined as $3h$, and the effective temperature gradient is approximated as $[T_1(x=0)-T_3(x=l_f)]/l_f$. Following the form of the Fourier's Law, thermal conductivity of SF-GF $K$ is obtained as

$$K=\dfrac{J_0 l_f}{3h\left[T_1(x=0)-T_3(x=l_f)\right]}=\dfrac{k_g t}{h\left[1+\dfrac{4e^{2\lambda l_f}-4}{\lambda l_f\left(e^{2\lambda l_f}-e^{\lambda l_f}+1\right)}\right]} \quad (6)$$

Here, $h$ is adopted as 0.34 nm based on literature [13], which is consistent with the commonly used monolayer graphene thickness $t$ [14]. Thus, $K$ can be simplified as

$$K = \frac{k_{\mathrm{g}}}{1 + \dfrac{4e^{2\lambda l_{\mathrm{f}}} - 4}{\lambda l_{\mathrm{f}}\left(e^{2\lambda l_{\mathrm{f}}} - e^{\lambda l_{\mathrm{f}}} + 1\right)}} \qquad (7)$$

Noticeably, $K$ shows a direct dependency on a dimensionless parameter $\lambda l_{\mathrm{f}}$, which combines synthetic effects of grafold geometry and graphene thermal properties.

## 3. Simulation methods

MD simulations has been widely used to study the thermal properties of nanomaterials [15-17]. In order to validate predictions by this analytical model, the temperature distribution and thermal conductivity of SF-GF are also calculated using MD simulation. The simulation model shown in Fig. 1e is based on the Reverse Non-Equilibrium Molecular Dynamics (RNEMD) method [18]. The hot region is placed at the center of simulation box while the cold regions are set at the both ends. Then the exchange of kinetic energy between atoms in hot and cold regions introduces a heat flow across the folded graphene structure from layer 1 to layer 3. The fold length $l_{\mathrm{f}}$ varies from around 15 to 100 nm. The width (*y*-direction) of models is kept as 5 nm. And the distance between two grafolds in the left and right parts of box is 0.7 nm, which is large enough to eliminate the effect of end-to-end adhesion on the heat transfer.

In this paper, MD simulations are conducted using the large-scale atomic/molecular massively parallel simulator (LAMMPS) [19]. The AIREBO potential is employed to describe the in-plane covalent bonds between carbon atoms, as well as the non-bonded van der Waals interactions between folded graphene layers [20]. This potential has been adopted by many simulation studies on the mechanical and thermal properties of carbon-based materials [21, 22]. The time step in all simulations is set as 0.5 fs. Initially, the atomic structure is optimized to equilibrium status by the conjugate gradient minimization method. Then the whole system is relaxed at 300 K under NPT ensemble for 200 ps, followed by the exchange of kinetic energy (every 200 timesteps) under NVE ensemble for more than 1 ns. The total heat flux $Q$ is calculated by the slope of the exchanged energy per unit cross-section area versus the simulation time curve. The graphene layers 1, 2 and 3 are separated into several small blocks, and the temperature of each block is calculated from MD simulations to form the temperature profiles $T_1(x)$, $T_2(x)$ and $T_3(x)$, respectively. After long enough simulation, the temperature profile of the system is stabilized, indicating the achievement of a stable heat transfer. Similar to the form in Eq. (6), the $K$ of simulation system is calculated as

$$K = \frac{Ql_f}{2\left[T_1(x=0) - T_3(x=l_f)\right]} \tag{8}$$

Noted that the factor 2 is attributed to the symmetry of the systems.

Moreover, as two essential input parameters for the calculation of $K$ using Eq. (7), values of $k_g$ and $\lambda l_f$ for the SF-GF with varying $l_f$ are also obtained by MD simulations. It should be noticed that $k_g$ is not a constant for simulation models with different $l_f$ since the in-plane transferring distance of heat flow (around $3l_f$) is smaller than the in-plane phonon mean free path of graphene around 775 nm [23]. In our analysis, $k_g$ of SF-GF models with $l_f$ from 15 to 100 nm is correspondingly taken from the thermal conductivity of another flat graphene models with lengths of $3l_f$ from 45 to 300 nm using the same RNEMD method. The atomic model for the calculation of $k_g$ is shown in **Supplementary Material**. Meanwhile, the value of $G$ is taken as $2.6\times10^7$ Wm$^{-2}$K$^{-1}$ based on a previous simulation result [15], which is insensitive to the in-plane size [13]. Thus, another input variable $\lambda l_f$ can be calculated. The results of $k_g$ and corresponding values of $\lambda l_f$ are listed in **Supplementary Material**. In addition, all the simulation cases show similar values of $J_0/t$ due to the same simulation settings, which is given as 62.4 Wm$^{-2}$ in the predictions of temperature profiles with Eq. (5).

**4. Results and discussions**

*4.1 Temperature profiles of SF-GF with different geometries of grafolds*

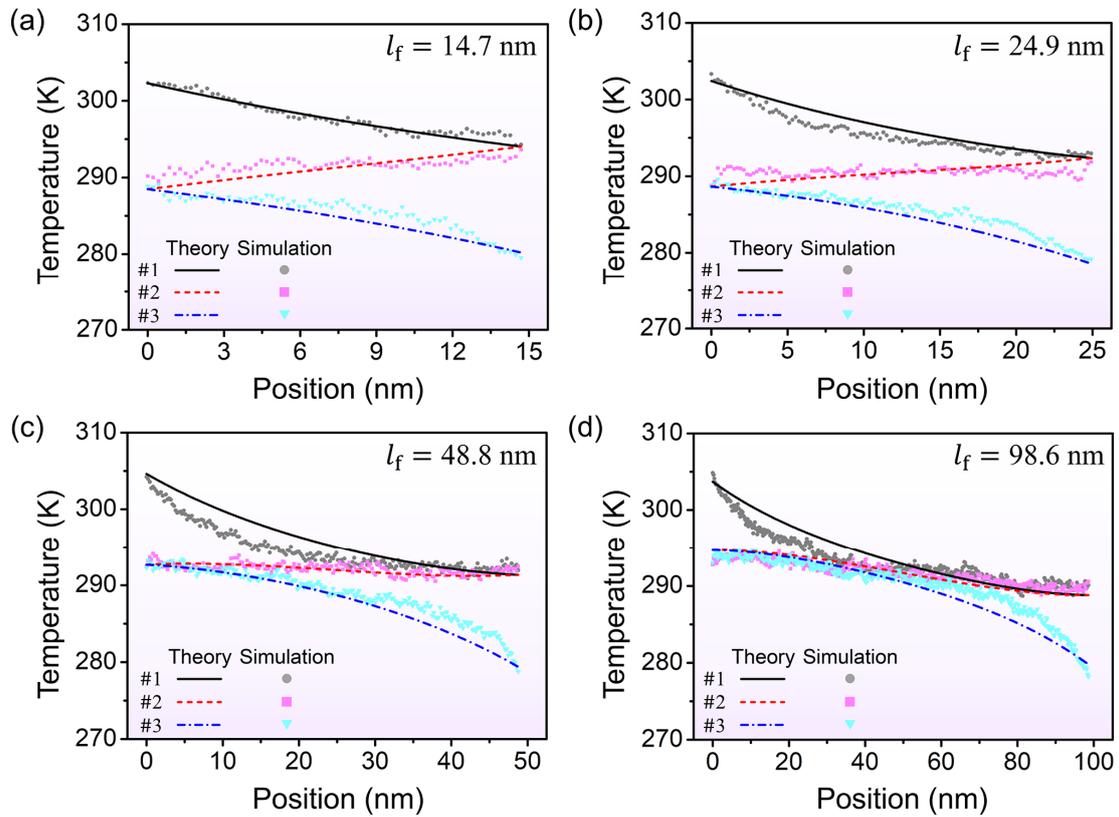

**Fig. 2.** Theoretical predictions and simulation results of temperature distributions in SF-GF with different fold lengths $l_f$. #1, #2 and #3 represent the graphene layer 1, 2 and 3, respectively.

The comparisons between the theoretical predictions and simulation results of temperature distributions are shown in Fig. 2. Clearly, temperature profiles derived by this model agree well with those output from MD simulations for SF-GF with different

$l_\mathrm{f}$, which provides a good validation for the description of heat transfer in SF-GF. It is interestingly found that $T_1$, $T_2$ and $T_3$ at the middle part of graphene layers becomes closer with increasing $l_\mathrm{f}$, even almost overlapped for the case with $l_\mathrm{f} = 98.6$ nm. In other words, for a SF-GF model with large $l_\mathrm{f}$, the interlayer heat transfer is mainly concentrated in the limit domains that are adjacent to the inflow and outflow ends of RVE, while it is significantly weakened at sufficiently large distances from the two ends. Such unique thermal behavior is reminiscent of the Saint-Venant principle applied in the uniaxial tension test, where the complicated contact forces between the sample and holder only affects the small area near the both ends of sample. It is also observed that the main trend of temperature distribution for layer 2 is reversed from downward to upward as $l_\mathrm{f}$ increases. The critical status, where $T_2$ is almost a constant, is achieved around $l_\mathrm{f} = 48.8$ nm, suggesting a nearly homothermal graphene layer no matter how much heat flux is introduced into SF-GF.

*4.2 Thermal conductivities of SF-GF with different geometries of grafolds*

The good agreement on temperature profile with MD simulation results suggests that this analytical model is promising as a more efficient tool to estimate thermal conductivity of SF-GF, especially for the applications in the practical graphene assemblies composed of graphene sheets mainly in several micrometers [24, 25]. For this case, MD simulation is time/resource-consuming for the predictions of thermal properties of SF-GF, while the analytical model can play its strength with more convenience. Under the length scale in micron order, the total in-plane transferring distance in SF-GF is usually far beyond the phonon mean free path of graphene, and

$k_g$ in Eq. (7) can be now regarded as a constant around 1000 Wm$^{-1}$K$^{-1}$ as the experimental measurement [26] instead of a length-dependent quantity. Then a normalized thermal conductivity can be introduced as $\overline{K} = K/k_g$ in the analysis for SF-GF in practical applications.

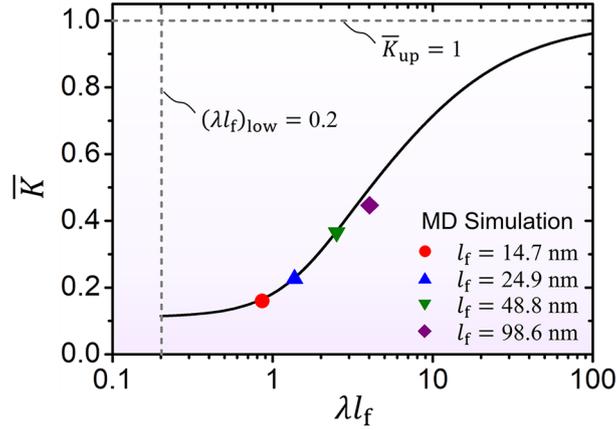

**Fig. 3.** Relationship between the normalized thermal conductivity of SF-GF $\overline{K}$ with the dimensionless parameter $\lambda l_f$. MD simulation results are plotted as data points for comparisons.

Fig. 3 shows the evolution of $\overline{K}$ with the dimensionless parameter $\lambda l_f$ obtained by this model. It should be mentioned here that $\lambda l_f$ has a lower bound around 0.2 based on the critical fold length $l_f^c$ for structural stability (See **Supplementary Material** for the structural stability analysis of SF-GF). In addition, the corresponding $\overline{K}$ and $\lambda l_f$ for the above MD simulation cases are also plotted as data points in Fig. 3. Our theoretical predictions of thermal conductivities for different $\lambda l_f$ are highly consistent with these simulation results, which confirms the accuracy of our analytical model again. From the $\overline{K}$ versus $\lambda l_f$ curve, thermal conductivities of SF-GF is always less than the intrinsic in-plane thermal property of graphene, since the heat transfer capability across graphene layers is much weaker than that within graphene planes. As $\lambda l_f$ increases, $\overline{K}$ goes up firstly. The rise of $\overline{K}$ is attributed to the expansion of folded area that provides more channels for interlayer heat transfer. Then it gradually converges to the upper bound $\overline{K} = 1$. It implies that the effect of folded microstructure on thermal conductivity vanishes for SF-GF with extremely large $\lambda l_f$. The effective manipulation of thermal conductivity by the folded structure can be only achieved for SF-GF with relatively small $\lambda l_f$.

*4.3 Evolution of thermal conductivity of SF-GF with stretch*

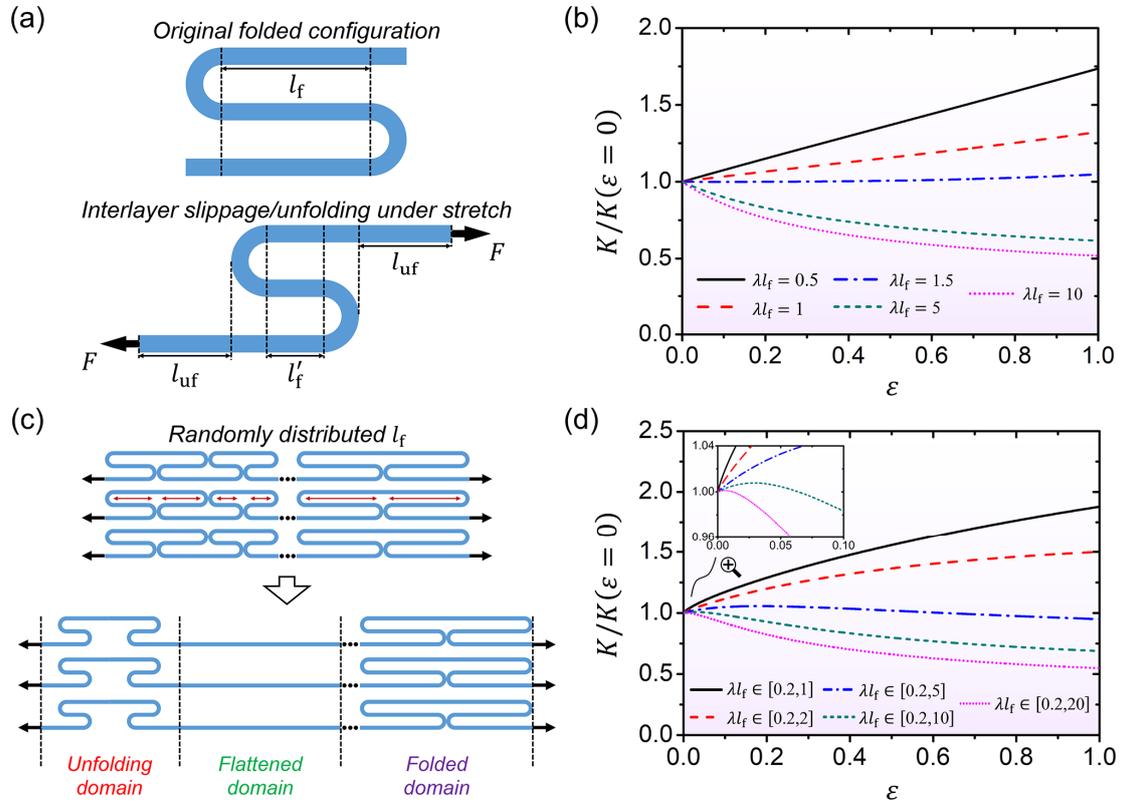

**Fig. 4.** (**a**) Schematic of the deformation in RVE under stretch. (**b**) Strain dependence of thermal conductivity of SF-GF. (**c**) Schematic of the deformation in SF-GF with randomly distributed fold lengths $l_f$ under stretch. (**d**) Strain dependence of thermal conductivity of SF-GF considering randomly distributed $l_f$.

The folded configuration of graphene can be unfolded and flattened under mechanical stretch [10], especially for the applications in strain sensors and other flexible electronics, which brings unavoidable change of thermal conductivity. Here, the analytical model is applied to investigate the evolution of $K$ under stretch for a SF-GF with original fold length $l_f$. It is well known that the interlayer van der Waals interaction is much weaker than the intralayer covalent bonds for graphene. That induces the structure failure initiating from interlayer slippage and causes negligible intralayer deformation under stretch [10]. For simplification, we only consider the effect of the deformation from the interlayer slippage on $K$ during stretch. Ignoring the length of joints between layers as well, the geometrical constraint requires that $2l_{uf} = 3(l_f - l'_f)$, where $l'_f$ and $2l_{uf}$ stand for the lengths of still folded part and unfolded part in the RVE after stretch, respectively, as illustrated in Fig. 4a. And the thickness of RVE is still defined as $3h$ before it is fully flattened. With the tensile strain in RVE as $\varepsilon = (2l_{uf} + l'_f - l_f)/l_f$, we can derive the function of $K$ with $\varepsilon$ as follows

$$K(\varepsilon) = \frac{2l_{uf} + l'_f}{3h\left(\dfrac{2l_{uf}}{k_g t} + \dfrac{l'_f}{3hK'}\right)} = \frac{(2+\varepsilon)k_g K'}{9\varepsilon K' + (2-\varepsilon)k_g} \qquad (9)$$

where $K'$ is the thermal conductivity of still folded part that can be also written as a function of $\varepsilon$ as

$$K'(\varepsilon) = K\big|_{l_f = l'_f} = \frac{k_g}{1 + \dfrac{8e^{(2-\varepsilon)\lambda l_f} - 8}{(2-\varepsilon)\lambda l_f \left[e^{(2-\varepsilon)\lambda l_f} - e^{(2-\varepsilon)\lambda l_f/2} + 1\right]}} \qquad (10)$$

It should be mentioned here that, the effect of strain on $k_g$ is beyond the scope of our consideration due to negligible intralayer deformation, although it is verified in many simulation and theoretical studies [27, 28]. Fig. 4b shows the evolutions of $K$ with $\varepsilon$ before SF-GF is completely flattened, where $K$ is normalized by $K(\varepsilon = 0)$. For the

SF-GF with $\lambda l_f$ less than 1.5, $K$ increases with tensile strain. However, it shows a decline with strain for the SF-GF with $\lambda l_f$ larger than 1.5. And $K$ is almost strain-insensitive with $\lambda l_f$ being around 1.5. The difference on the geometry of grafold not only determines the thermal conductivity of SF-GF but also reverses its monotonicity with the increase of tensile strain.

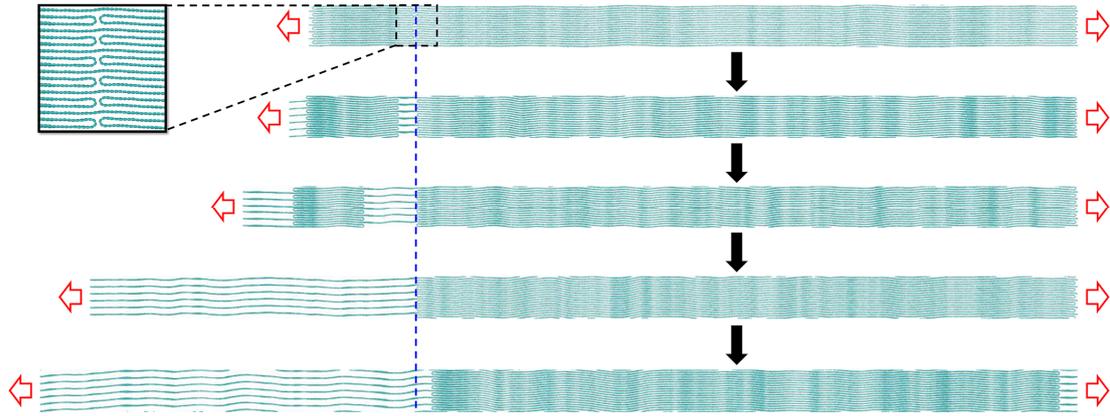

**Fig. 5.** Snapshots of MD simulations for SF-GF model composed of grafolds with non-uniform fold lengths $l_f$ under stretch.

In fact, $l_f$ varies in practical graphene assemblies, instead of following the uniform assumption as our idealized model. In order to show the structural irregularity, we further develop a SF-GF model that contains a series of grafolds with dissimilar fold lengths $l_f$, as shown in Fig. 4c. Previous study found that the grafold with small $l_f$ shows less failure stress than that with large $l_f$ [10]. And our MD simulations confirm that the grafold with small $l_f$ would be unfolded and flattened preferentially during stretch, as the snapshots of MD simulations shown in Fig. 5. It is noticed that the grafold with large $l_f$ begins to be unfolded until the one with small $l_f$ is completely flattened. During the unfolding process of the grafold with small $l_f$, there is almost no deformation caused by stretch in another grafold with large $l_f$. An assumption is reasonably proposed in our analysis that grafolds with dissimilar $l_f$ in SF-GF will be unfolded one by one following the sequence of increasing $l_f$, with the evidence provided by this simulation result. Thus, the grafolds in the SF-GF model shows three different stages during stretch as shown in Fig. 4c, which divide the overall structure of SF-GF into still folded, unfolding and flattened domains.

Consider a SF-GF model composed of grafolds with dissimilar fold lengths $l_f$ randomly distributed in a given range, and the total number of grafolds $N$ is set at 1000. According to the assumption, there is only one unfolding grafold in every moment of stretch for idealization. With the progress of stretch, more grafolds are unfolded, which updates the numbers of grafolds within still folded and flattened domains. Obviously, the thermal conductivity of flattened domains exactly equals to $k_g$. In addition, thermal conductivities of still folded and unfolding domains have been already obtained by Eqs. (7) and (9-10), respectively. Thus, the evolution of $K$ with the $\varepsilon$ for the SF-GF with random $l_f$ can be derived as

$$K(\varepsilon) = \frac{\sum_{i=1}^{p} l_f^{\text{folded}}\big|_{(i)} + (1+\varepsilon_{\text{local}})l_f^{\text{unfolding}} + \sum_{j=1}^{q} 3 l_f^{\text{flattened}}\big|_{(j)}}{\sum_{i=1}^{p} \dfrac{l_f^{\text{folded}}\big|_{(i)}}{K_{\text{folded}}\big|_{(i)}} + \dfrac{(1+\varepsilon_{\text{local}})l_f^{\text{unfolding}}}{K_{\text{unfolding}}} + \sum_{j=1}^{p} \dfrac{l_f^{\text{flattened}}\big|_{(j)}}{k_g}} \qquad (11)$$

where $l_f^{\text{folded}}\big|_{(i)}$, $l_f^{\text{unfolding}}\big|_{(j)}$ and $l_f^{\text{flattened}}$ are the original fold lengths of grafolds belonging to still folded, unfolding and flattened domains before stretch, respectively. $p$ and $q$ represent the numbers of grafolds from still folded and flattened domains, respectively, which are updated during the stretch process. And $\varepsilon_{\text{local}}$ in Eq. (11) stands for the local tensile strain of the unfolding grafold that is equivalent to the strain

in Eqs. (9) and (10), which contributes to the total strain $\varepsilon$ as

$$\varepsilon = \frac{\sum_{i=1}^{p} 3 l_f^{\text{flattened}}\big|_{(i)} + (1+\varepsilon_{\text{local}}) l_f^{\text{unfolding}} + \sum_{j=1}^{q} l_f^{\text{folded}}\big|_{(j)}}{\sum_{k=1}^{N} l_f\big|_{(k)}} - 1 \quad (12)$$

with the approximation of original total length of SF-GF as $\sum_{k=1}^{N} l_f|_{(k)}$. Combining Eq. (11) and (12), the theoretical results before SF-GF is fully flattened are shown in Fig. 4d. For $\lambda l_f$ within narrow ranges such as $\lambda l_f \in [0.2, 1]$ and $\lambda l_f \in [0.2, 2]$, $K$ rises monotonically with increasing $\varepsilon$. However, as the range of $\lambda l_f$ is enlarged, $K$ drops at a large $\varepsilon$ and exhibits a non-monotonic variation with $\varepsilon$, namely, increases firstly then decreases with $\varepsilon$. The initial increment of $K$ is suppressed with the broadening range of $\lambda l_f$.

In many regards, thermal conduction is analogous to electrical conduction [29] that is easier to be measured. Interestingly, some experimental studies found that the electrical conductance of graphene strain sensor with wrinkles increases initially then drops during uniaxial tension as well [30]. It is noted that the predicted non-monotonic evolution of $K$ with tensile strain plotted in Fig. 4d coincides with such experimental phenomenon. Considering that it is very likely to form grafolds by collapsed graphene wrinkles [6], the analytical model proposed in this paper would provide an possible interpretation on the mechanisms behind this particular phenomenon in the applications of graphene-based strain sensor.

## 5. conclusions

In summary, an analytical model is developed for the efficient prediction of temperature profile and thermal conductivity of SF-GF material with good validations using MD simulations. The relationship between the folded microstructure and overall thermal properties of SF-GF is demonstrated in this work. Using this model, the evolution of thermal conductivity of SF-GF with the unfolding deformation during stretch is also investigated. In addition, the effect of structural irregularity on the thermal properties of SF-GF is further considered in our analysis. It is found that the thermal conductivity of SF-GF can show a non-monotonic dependence on the tensile strain. Such behavior fits the analogous experimental phenomenon observed in graphene-based strain sensor. The model not only benefits the manipulation on the thermal properties of SF-GF but also gives inspirations to the in-depth understanding of some related experimental phenomenon in applications of graphene-based devices.

**Declaration of Competing Interest**

None

**Acknowledgment**

This work was supported by National Natural Science Foundation of China (Grant No. 11972226). The computational work was supported by the HPC Center of Shanghai Jiao Tong University.